\documentclass[runningheads]{svmult}

\usepackage{makeidx}  
\usepackage{graphicx} 
\usepackage{subeqnar} 
\usepackage{multicol} 
\usepackage{cropmark} 
\usepackage{physprbb} 
\makeindex            

\def\gtsima{$\; \buildrel > \over \sim \;$}
\def\ltsima{$\; \buildrel < \over \sim \;$}
\def\gsim{\lower.5ex\hbox{\gtsima}}
\def\lsim{\lower.5ex\hbox{\ltsima}}
\begin{document}
\title*{Probing the evolution of massive galaxies with the K20
survey\thanks{The collaborators in the K20 survey include T. Broadhurst
(HUJ), S. Cristiani (ECF \& Trieste), S. D'Odorico (ESO), E. Daddi 
(Firenze \& ESO), A. Fontana (Roma), E. Giallongo (Roma), R. Gilmozzi
(ESO), N. Menci (Roma), M. Mignoli (Bologna), F. Poli (Roma), L.
Pozzetti (Bologna), S. Randich (Arcetri), A. Renzini (ESO), P. Saracco 
(Milano), J. Vernet (Arcetri), G. Zamorani (Bologna)}
}
%
%
%
%
\titlerunning{Massive galaxies in the K20 survey}
%
\author{Andrea Cimatti}
\authorrunning{Andrea Cimatti}
%
%
\institute{Osservatorio Astrofisico di Arcetri, Largo E. Fermi 5,
I-50125, Firenze, Italy}

\maketitle              

\begin{abstract}
The motivations and the status of the K20 survey are presented.
The first results on the evolution of massive galaxies and the 
comparison with the predictions of the currently competing scenarios 
of galaxy formation are also discussed.
\end{abstract}

\section{Introduction}

Understanding the evolution of massive galaxies (e.g. 
M$_{stellar}$\gsim$10^{11}$ M$_{\odot}$) is important
because to constrain the 
different scenarios of structure and galaxy formation.
In particular, the question on the formation of the present-day 
massive spheroidals is still one of the most debated issues of 
galaxy evolution (see \cite{renz} for a review).
In one scenario, massive spheroidals are formed at early cosmological 
epochs (e.g. $z>3$) through a short and intense episode of star 
formation (with $SFR\sim 100-1000$ M$_{\odot}$yr$^{-1}$), followed 
by a passive evolution (or pure luminosity evolution, PLE) of the 
stellar population to nowadays.
In marked contrast, the hierarchical scenarios predict that massive 
spheroidals are the product of rather recent merging of pre-existing
disk galaxies taking place mostly at lower redshifts and with moderate
star formation rates \cite{k96,bau}. In hierarchical merging scenarios, 
fully assembled massive field spheroidals with M$_{stellar}$\gsim$10^{11}$
M$_{\odot}$ at $z$\gsim 1 are very rare objects \cite {k98} (see also Baugh, 
this volume). 

From an observational point of view, a direct way to test the above
scenarios is to study the evolution of massive galaxies by means
of spectroscopic surveys of field galaxies selected in the $K$-band
\cite{bro,cowie,cohen,stern}.  
Since the near-IR light is a good tracer of the galaxy mass\cite{gava,k98}, 
$K$-band imaging allows to select massive 
galaxies at high-$z$. A galaxy with a stellar 
mass of about $10^{11}$ M$_{\odot}$ is expected to have $18<K<20$ for 
$1<z<2$ \cite{k98}, thus implying that moderately deep $K$-band surveys 
can efficiently select massive galaxies in that redshift range. Deep 
spectroscopy with 8-10m class telescopes can then be used to search for
massive systems and to constrain their redshift distribution.

\section{The K20 survey}

\subsection{The observations and the database}

In order to investigate the evolution of massive galaxies and to
constrain the currently competing galaxy formation scenarios, we
started in 1999 a project that was called ``K20 survey''. For such
a project, 17 nights were allocated to our team in the context  
of an ESO VLT Large Program distributed over a period of two years 
(1999-2000) (see also {\tt http://www.arcetri.astro.it/$\sim$k20/}).

The prime aim of such a survey is to derive the redshift distribution of
about 550 $K$-selected objects with $Ks\leq20$. The targets were 
selected from a 32.2 arcmin$^2$ area of the Chandra Deep Field South 
(CDFS; \cite{giacconi}) using the images from the ESO Imaging Survey 
public database (EIS; {\tt http://www.eso.org/ \\ science/eis/}; the 
raw $Ks$-band images were reduced and calibrated by our group ), and 
from a 19.8 arcmin$^{2}$ field centered at 0055-269 using NTT+SOFI 
$Ks$-band imaging (Fontana et al. in preparation). 

Optical multi-object spectroscopy was made with the ESO VLT UT1 and UT2
equipped with FORS1 (October-November 1999) and FORS2 (November 2000) 
during 0.5$^{\prime\prime}$-1.5$^{\prime\prime}$ seeing conditions and 
with 0.7$^{\prime\prime}$-1.2$^{\prime\prime}$ wide slits depending on 
the seeing. The grisms 150I, 200I, 300I were used with typical integration 
times of 1-3 hours. Dithering of the targets along the slits between 
two fixed positions was made for most observations in order to 
efficiently remove the CCD fringing and the strong OH sky lines at 
$\lambda_{obs}>7000$~\AA. The spectra were calibrated using standard 
spectrophotometric stars, dereddened for atmospheric extinction, corrected 
for telluric absorptions and scaled to the total $R$-band magnitudes. 
A small fraction of the K20 sample
was observed with near-IR spectroscopy using the VLT UT1+ISAAC in order
to derive the redshifts of the galaxies which were too faint for
optical spectroscopy and/or expected to be in a redshift range
for which no strong features fall in the observed optical 
spectral region (e.g. $1.5<z<2.0$). However, due to the lack of a 
multi-object spectroscopy mode in ISAAC, it turned out very hard 
and inefficient to obtain redshifts of faint galaxies in this manner, 
which was successful only for a few galaxies at $1.3<z<1.9$ with 
strong H$\alpha$ in emission.

In addition to spectroscopy, $UBVRIzJKs$ imaging is also available 
for both fields, thus providing the possibility to estimate photometric 
redshifts for all the objects in the K20 sample, to ``calibrate'' them 
through a comparison with the spectroscopic redshifts and to assign 
reliable photometric redshifts to the objects for which it was not 
possible to derive the spectroscopic $z$ (see Fontana et al., these
proceedings).

The spectroscopic observations were completed in December 2000. 
The spectral analysis was done by means of automatic software (IRAF: 
{\tt rvidlines} and {\tt xcsao}) and through visual inspection of the 
1D and 2D spectra.
Because of four nights lost due to the bad weather, only 94\% of the 
sample with $Ks<20$ could be observed. The efficiencies in deriving
the specroscopic redshifts for the observed targets was high:
$N_{identified}/N_{observed}$=95\%, 93\%, 91\% for $Ks<19.0,\ 19.5,
\ 20.0$ respectively. The overall spectroscopic redshift completeness 
is still rather
high, with $N_{identified}/N_{total}$=93\%, 91\%, 85\% for $Ks<19.0, \
19.5, \ 20.0$ respectively (where $N_{total}=N_{observed}+N_{unobserved}$). 
The size of the sample, the spectroscopic redshift completeness,
and the availability of tested and reliable photometric redshifts make 
the K20 sample one of the largest and most complete database to study the 
evolution of $K$-selected galaxies available to date.

\subsection{The scientific aims}

Kauffmann \& Charlot (1998) estimated that $\sim 60\%$ and $\sim 10\%$
of the galaxies in a $K<20$ sample are expected to be at $z>1$, 
respectively in a PLE and in a standard CDM hierarchical merging model 
(cf. their Fig. 4). Such a large difference was in fact one of the main 
motivations of our original proposal to undertake a redshift survey for
all objects down to $K<20$. However, more recent models consistently
show that the difference between the predictions is less extreme than 
in the KC98 realization (Menci et al., Pozzetti et al., Somerville et 
al., in preparation). Part of the effect is due to the now favored
$\Lambda$CDM cosmology which pushes most of the merging activity at 
earlier times compared to $\tau$CDM and SCDM models, and therefore 
get closer to the PLE case. Moreover, a different tuning of the 
star-formation algorithms (to accomodate for more star formation
at high $z$) also reduces the differences between the two scenarios.
Our database is currently being used to perform a stringent comparison
between the observed redshift distribution and the ones predicted by
the most recent models of galaxy formation (Cimatti et al., in
preparation).

Besides the main goal described here above, our unique database is also
being used to address other important questions on galaxy evolution:
{\it (1)} the evolution of the Luminosity and Mass Functions (see
Pozzetti et al., this volume), {\it (2)} the evolution of ellipticals, 
{\it (3)} the fraction of dusty starbursts and high-$z$ ellipticals 
in the ERO population, {\it (4)} the evolution of galaxy clustering 
(see Daddi et al., this volume), {\it (5)} the spectral properties 
of a large number of galaxies and their evolution as a function of 
redshift, {\it (6)} the volume star formation density 
using different indicators, {\it (7)} the fraction of AGN in $K$-selected 
samples, {\it (8)} the brown dwarf population at high Galactic latitude.

\section{First results on Extremely Red Objects}

\begin{figure}
\centering
\includegraphics[width=1.0\textwidth, height=12cm]{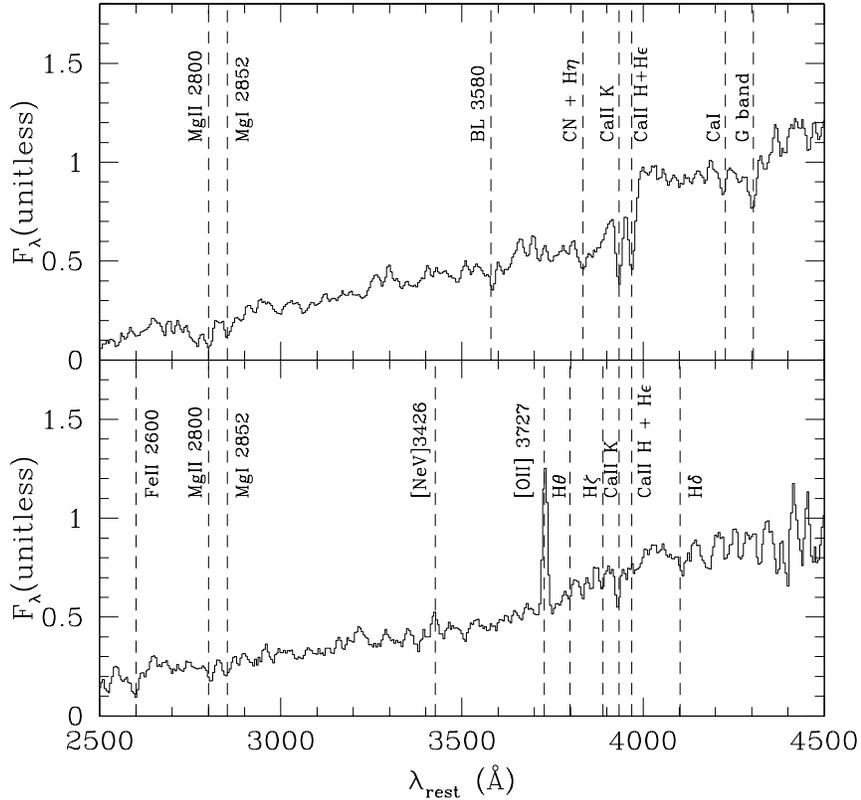}
\noindent
\caption{
The average rest-frame spectra (smoothed with a 3 pixel boxcar) of 
old passively evolving (top; $z_{mean}=1.000$) and dusty 
star-forming EROs (bottom; $z_{mean}=1.096$) with $Ks\leq20$.
}
\label{eps1}
\end{figure}

A fraction of the galaxies selected in the near-infrared show very
red colors (e.g. $R-K>5$). Such galaxies are known as Extremely
Red Objects (EROs), and the most recent surveys demonstrated that 
they form a substantial field population \cite{tho,daddi1,mc}.
Since their colors are consistent with being either old passively
evolving galaxies or dusty starbursts or AGN, it is therefore of prime
importance to determine their nature in order to exploit the stringent
constraints that EROs can place on the galaxy formation scenarios.
In this section we adopt a cosmology with $H_0=70$ km s$^{-1}$
Mpc$^{-1}$, $\Omega_m=0.3$ and $\Omega_{\Lambda}=0.7$.

\begin{figure}
\centering
\includegraphics[width=1.0\textwidth, height=12cm]{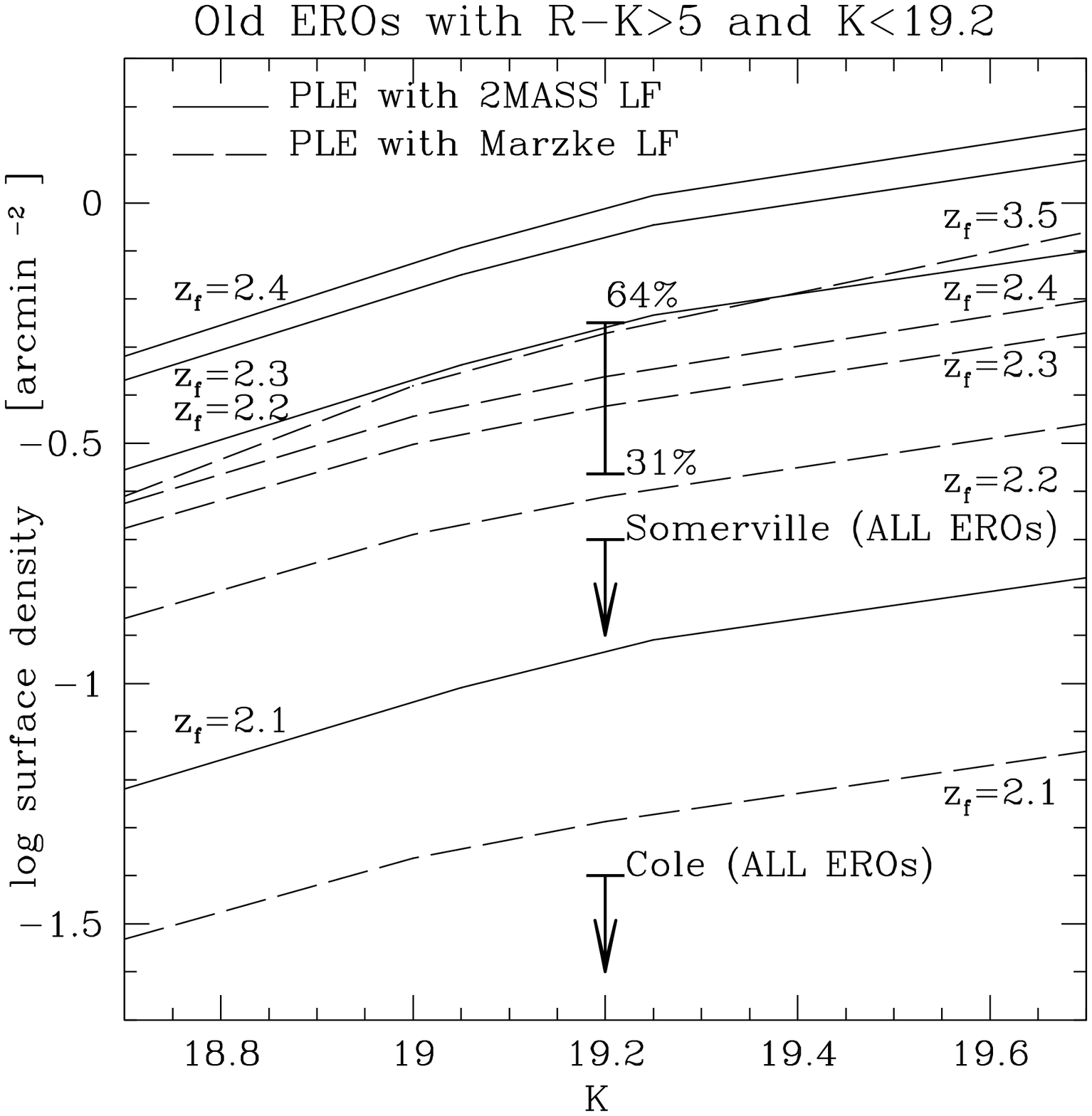}
\noindent
\caption{
Comparison between the observed density of old EROs with the predictions
of PLE models (see \cite{c01} for more details). The range of 
observed density of old EROs with $Ks<19.2$ is shown with a vertical bar 
ranging from the minimum observed density (32\% of the total density 
of EROs observed in the K20 survey) and the maximum possible density 
(64\% of the total density of EROs assuming that all the spectroscopically 
unidentified EROs are old passive systems). The dashed and solid lines 
show the predictions of PLE models adopting respectively the local 
luminosity function of ellipticals \cite{marzke} and of 
early-type galaxies (2MASS, \cite{kocha}), and using Bruzual 
\& Charlot spectral synthesis models with solar metallicity, Salpeter 
IMF, $e$-folding time of the star formation $\tau$=0.3 Gyr, and a set 
of formation redshifts ($z_f$). See also \cite{daddi2}
for more details on PLE models. The arrows indicate the predicted
densities of EROs in two recent hierarchical merging models (the
Somerville et al. model shown in Firth et al. 2001, and the Cole et al.
2000 model presented in Smith et al. 2001). Since such models predict the 
total density of {\it all} EROs (i.e. old + dusty star-forming) they 
actually represent upper limits to the predicted density of old 
passive systems.
}
\label{eps1}
\end{figure}

From our total $Ks<20$ sample, 78 EROs with $R-Ks\geq5.0$ were extracted.
About 70\% of the EROs with $R-Ks\geq5.0$ and $Ks\leq19.2$ was 
spectroscopically
identified with {\it old} and {\it dusty star-forming} galaxies
at $0.7<z<1.5$ \cite{c01}. The two classes are about equally 
populated and for each of them we derived the average spectrum (Fig. 1).

Old EROs have an average spectrum consistent with being old passively 
evolving ellipticals, and it can be well reproduced with Bruzual \& 
Charlot (2000) spectral synthesis models with no dust extinction and 
ages \gsim 3 Gyr if $Z=Z_{\odot}$, thus implying an average formation 
redshift $z_f$\gsim 2.4 for such a metallicity. 

The average spectrum of star-forming EROs can be reproduced only 
if a substantial dust extinction is introduced, typically in the range 
of $0.5<E(B-V)<1$. Their star formation rates, corrected for the 
average reddening, suggest a significant contribution ($>$20\%) 
of EROs to the cosmic star-formation density at z$\sim$1. However,
the detection
of [NeV]$\lambda$3426 emission suggests that a fraction of 
star-forming EROs also host some AGN activity obscured by dust.

Since old EROs have spectra consistent with being passively evolving
ellipticals, we compared their density with different model predictions
(see Fig. 2). The main result is that the density of old EROs observed
in our survey is strongly underpredicted by the current hierarchical 
merging models (a factor of $\sim$4--10 for the models presented by
\cite{firth} and \cite{smith} respectively). On the other
hand, PLE models with a reasonable choice of input parameters predict
surface and comoving densities in agreement with the observations
and imply that massive spheroidals formed at $z_f>2-3$ (see \cite{c01} 
for more details). 

Since the luminosities and the stellar masses of the observed old EROs 
are in the range of 0.5-4$L^{\ast}$ and 1-6$\times 10^{11}$ M$_{\odot}$, 
this means that fully assembled massive systems were already in place 
up to $z\sim 1.3$. 

The existence of such galaxies with a comoving density 
of \gsim$2\times 10^{-4}$ Mpc$^{-3}$ at $z\sim 1$ (\cite{c01}) is in 
contrast with the $\Lambda$CDM hierarchical model of \cite{cole}, where 
the comoving density of {\it all galaxy types} with M$>10^{11}$ M$_{\odot}$ 
is about one order of magnitude lower, whereas the difference 
is less dramatic with the model of \cite{k99} (see Fig. 1 of \cite{benson}).

\clearpage
\addcontentsline{toc}{section}{Index}
\flushbottom
\printindex

\end{document}